\documentclass[10pt,journal]{IEEEtran}

\usepackage{graphicx,changepage}
\usepackage{float}
\usepackage{subfigure}
\usepackage{scalerel}
\usepackage{amsmath}
\usepackage{xcolor}
\usepackage{cite}
\usepackage{array}
\usepackage{pifont}
\usepackage{fixltx2e}
\usepackage{relsize}

\ifCLASSINFOpdf
\else
\fi

\begin{document}
\title{Superconducting Quantum Amplifier-Integrator in Ultra-High Speed Continuous-time $\Delta$-$\Sigma$ Converter}
\author{\IEEEauthorblockN{Debopam Banerjee}\\
\IEEEauthorblockA{Analog Devices, India\\}}

\markboth{IEEE Journal Name,~Vol.~**, No.~**, May~2020}%
{Shell \MakeLowercase{\textit{et al.}}: Bare Advanced Demo of IEEEtran.cls for IEEE Computer Society Journals}


\maketitle

\begin{abstract}
The current semiconductor research is increasingly focussing on either attaining higher speeds (Gigabits/s) or higher linearity (harmonic distortion in dB) or sometimes both of them. Applications for such technologies range from consumer to industrial to healthcare and to military. Typically such circuits are fabricated in today's low-voltage CMOS processes using Silicon and in few cases BJT-CMOS combined processes using Gallium-Arsenide or Indium-Phosphide. These technology nodes face a plethora of problems like reduction of dynamic range of the circuit due to mismatch, distortion, noise, thermal and electromigration issues due to excessive current densities with shrinking device geometries, off-state leakage currents, etc. Compounding these problems is the issue with lower achievable gain from an amplifier which often gets limited due to lower supply voltages in such technology nodes. Slowly circuit techniques like chopping, cascoding, cascading and calibration are nearing their limits. In this paper we present a radically different approach to our regular analog design building blocks using macroscopic quantum effects which have hitherto not found favour with the design community. We will solely focus on the effect of superconductivity and adopting its macroscopic phenomena to amplifiers, integrators and comparators. Using staggered superconductors we can achieve a gain which depends only on physical quantum constants and remains invariant under process, temperature, supply, interference, etc. This robustness of gain in an amplifier goes a long way in attaining higher linearity. The comparator can resolve a minimum of 2.07fT magnetic flux but when embedded inside the $\Delta$-$\Sigma$-loop can typically attain 100 times smaller resolution pushing the boundaries of sensing.
\end{abstract}


\IEEEpeerreviewmaketitle

\setlength{\parskip}{1em}
\makeatletter
\newcommand*{\rom}[1]{\expandafter\@slowromancap\romannumeral #1@}
\makeatother

\section{Introduction}
Around the the turn of the twentieth century, there were observations made from various experiments that challenged the accepted classical notion of particles. We have tried to leverage some of these Quantum concepts which manifest macroscopically and adapt them to our everyday circuits like amplifiers, integrators, ADCs, etc. A case in point being superconductivity which exhibits macroscopic quantum effects like "Meissner effect" and "Magnetic flux quantization" among others. \par

Out of the total numbers of integrated circuits fabricated all around the world, a overwhelming majority of them are using silicon (Si) as their base material. A few high-speed circuits class sometimes employ Gallium-Arsenide (GaAs) as the wafer material while a minuscule fraction uses other exotic \rom{3}-\rom{5} compounds like InP, GaN, etc. These are plagued with issues like -:
\begin{itemize}
  \item Mismatch amongst transistors in a particular chip and mismatch in same transistor from chip to chip. This reduces the achievable linearity of the circuit. Techniques to reduce this in the past have involved methods like chopping whereby the incoming signal is shifted to higher frequencies, processed, and downconverted back. But as evident, this will not work as our demand for higher and higher bandwidths keep increasing the speed demanded from these devices would also increase exponentially.
  \item Parasitic capacitances arising due to the self junction capacitance from the device, from routing the signals over the chip, and from the load that this device is driving. This also directly affects the operating speed of the device and in some cases causes linearity issues. Historically designers have got around this problem by burning more and more current in the device. But as we chase higher compaction and miniaturization the current density increases dramatically leading to gross failures from electromigration, device reliability, accelerated ageing of the transistor, etc.
  \item Another big headache that designers of such circuits have to deal with is the issue of mismatch from circuit to circuit within the same chip e.g.-: between current sources, between amplifiers, between voltage levels, etc. To get around these issues, people have come up with techniques like Data Weighted Averaging (DWA), Dynamic Element Matching (DEM), Shuffling and Dithering which basically randomizes the harmonic tones into white noise and then shape out the excess noise from the bandwidth of interest. This method though very effective in low-speed applications, has very little success in higher speeds and very high resolution applications. Moreover this introduces significant power, area and design complexity penalty.
\end{itemize}

While the above discussion shows the limits stemming from devices, we will briefly discuss the limits caused by circuits employing those devices. The issue of finite gain of an amplifier is by far the most limiting cause in analog circuits. This is primarily due to the lower geometries of device technology nodes and lower supplies. Here too in the past people have employed techniques like multi-stage gains, cascoding in higher supply voltages, etc. But both these methods cause significant bandwidth reduction due to multiple poles and cannot be used for really high-speed and high-linearity applications. Moreover this also requires expensive and complex compensation schemes for stabilization.

A $\Delta \Sigma$-ADC can very crudely be said to work on the principle of time averaging the outputs from a crude comparator and feeding them back with the input after proper amplifications to make it look like a very precise comparator. So even using two voltage levels like ``+1'' and ``-1'' we can represent an analog input like ``0.31415'' with good enough precision. As will be discussed in later sections, the proposed comparator in this paper has a resolution of 2.07fT (femto Tesla). So employing this comparator in our proposed $\Delta \Sigma$-ADC loop, we can theoretically detect and reliably measure much finer fields which would have been practically impossible using the existing conventional electronic circuits. The implications from this and the resulting applications that can be envisaged for such state-of-the-art solutions is wide and far-reaching. These can range from 
\begin{itemize}
  \item high precision scientific measurements like those required in particle colliders, high-energy experimentations, molecular research
  \item industrial measurements and control systems like those required in industrial automation, power-plants, equipment monitoring, control systems
  \item exploration and mapping
  \item communication systems
  \item diagnostics, healthcare and medicine, tomographic measurements, etc
  \item astronomical observations and space explorations
  \item applications in military and aerospace
\end{itemize}
With new promising research showing the existence of room temperature superconductivity, a lot of the problems associated with cryogenic setup, robustness and scalability can be addressed in the coming years.

The paper is arranged in six sections. In section-\rom{2} we will give a brief overview of an example circuit which we are trying to replace with this proposed approach and methods. Post that we will briefly compare the pros and cons of the proposed method over existing ones. In section-\rom{3} we will try to explain in simplest possible terms how it works and when it might not. In section-\rom{4} we will delve into detailed mathematical modelling of the proposed circuits and proceed to derive the various controlling fields like magnetic($\vec{B}$), electric($\vec{E}$) and supercurrent density($\vec{J^*_s}$). In section-\rom{5} we will look at the equations governing the Superconductor-Metal-Superconductor(SNS) junction during transition times when we turn on/off the controlling coils. Finally in section-\rom{6} we will have a limited overview of the Bose-Einstein theory central to the working of bosons as the Cooper-pairs in superconductors are practically Bosons. This will help us in understanding the fundamental limitations of these proposed devices.

\section{Proposed Circuits and Devices}
In this section we will first introduce the variety of circuits that we are trying to implement with our proposed technique. In the following figure we have a classical $\Delta$-$\Sigma$ modulator Analog to Digital Converter showing the blocks it is made up of $\rightarrow$ opamps, passives like capacitors and resistors, comparator, feedback elements comprising either current-steering DACs or resistive-DACs depending upon the application needs.\par
\begin{figure*}[!t]
\centering
\setlength\fboxsep{0pt}
\setlength\fboxrule{0.25pt}
\fbox{\includegraphics[width=\textwidth]{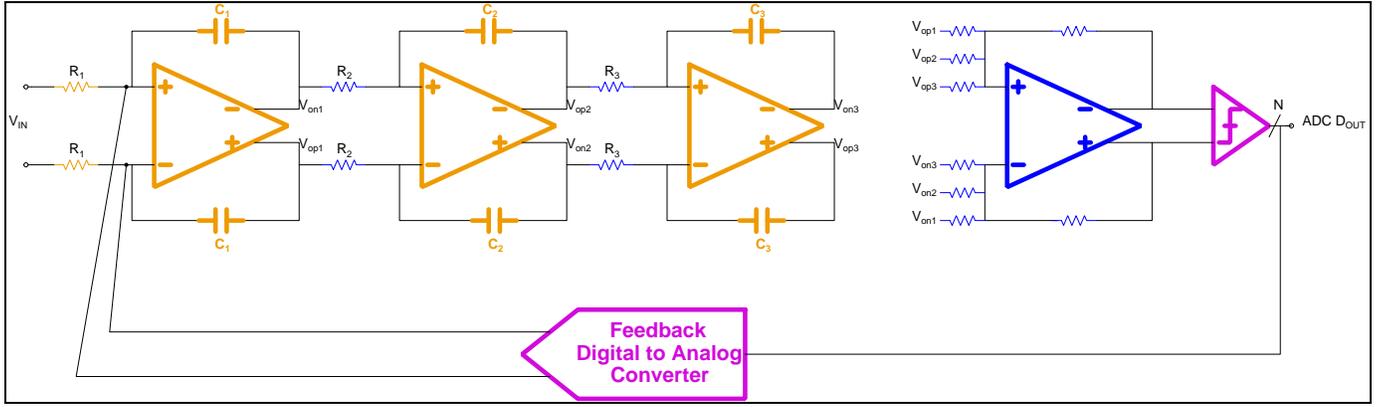}}
\caption{A classical $\Delta$-$\Sigma$ Modulator ADC showing the basic building blocks.}
\label{classical_DSM}
\end{figure*}

If the input varies really slow compared to the clocking speed of this ADC, the "average of the output" will track the "average of the input". The quality of this tracking is primarily dependant on the collective gain of the loop-filter preceding the comparator, the resolution of the feedback signal and the speed with which the ADC operates.The blocks highlighted in Fig. \ref{classical_DSM} are those which we are trying to replace with our proposed set of circuits. Typically the same $\Delta$$\Sigma$-ADC can be implemented via a feedforward or a feedback configuration. Here we will stick to a feedforward (CIFF) configuration to compare the implementations. 

The macroscopic superconducting phenomena that we are interested in harnessing are best captured by Meissner effect and trapping of external magnetic field when a material changes phase from normal to superconduting. This change can be brought about in 3 independent ways -:
\begin{itemize}
  \item Increasing the current or supercurrent density beyond the critical limit $J_C$
  \item Increasing the magnetic field strength beyond $H_C$ in type-\rom{1} superconductors or beyond $H_{C2}$ in type-\rom{2} superconductors
  \item Increasing the tempetature above the critical temperature $T_C$ for the superconductor
\end{itemize}

In our proposed methods we will be using the $2^nd$ point to selectively bring a superconductor in and out of it superconductive phase. The other effect that we will be using, the Meissner effect, is basically a phenomenological observation wherein if the material in the presence of an external magnetic flux field is cooled below $T_C$ then it expells the entire magnetic field from within its enclosed volume and acts as a perfect diamagnetic. This effect, as we will see later in the paper, is a more fundamental property and defining feature of superconductors than the oft quoted "zero resistivity". These are best explained in the following Fig. \ref{flux_trap}

To better understand why we have the flux stored we beign from Maxwell-Faraday equation in both differential and integral forms,
\begin{equation}
\nabla \times \vec{E} = -\frac{\partial{\vec B}}{\partial t}
\end{equation}
\begin{equation}
\oint_C \vec{E} \,dl = - \frac{\partial}{\partial t} \int_S \vec{B} \,dS
\end{equation}
Let us now consider the contour shown by dotted pink contour shown at the top of the cylinder. As seen clearly, this lies within the normal ohmic material in case-1. Evaluating the $2^{nd}$ equation above along this contour we conclude that the flux threading the 2-D surface enclosed by this contour must remain constant with time. For a superconductor due to Meissner effect, the total magnetic flux inside a superconducting cylinder is zero. Thus the magnetic field in the free space must remain constant and the integral on the left-hand side evaluates to zero. After the matierial moves to superconducting phase, an induced current of superelectrons or Cooper-pairs expel the flux inside the material. However an equal and opposite current of superelectrons flow in the opposite direction at the inner surface of the cylinder to keep the magnetic field constant. Now finally in case-3 when the magnetic field is turned off then the supercurrent on the outer surface ceases to flow but by the same logic as before, the supercurrent on the inner surface flows and the flux density originally present externally is now trapped.\par
This trapped current has been verified experimentally to have a decay constant in tens of years and thus for all practical purposes can be considered constant in our applications where each clock cycle might last not more than a few 100$\mu$s. Now that we have a clear way of converting the input signal, either voltage or current, into a magnetic field and process it we can move forward to our original plan of somehow amplifying or integrating it. This will solve the basic functionality of the saffron coloured blocks to build a $\Delta$-$\Sigma$ ADC.\par
Comparing the pros and cons of the aforementioned method, we notice without involved calculations the following -:
\begin{itemize}
  \item We can have a circuit that it totally devoid of process and voltage variations as compared to finer geometry CMOS processes.
  \item The minimum field that can be generated and stored is of the order of 2.07fT(femto Tesla). To get ourselves and idea of how fine this distinction is in terms of physical parameters, let us consider an example. Suppose we fabricate a solenoid with air filled core having 1 turn per 1nm, which is a modest number. For $\vec{B}$ to assume such low fields, we will need to send a current of 1pA through the coil woundings to generate a field small and stable enough to match that resolved in a SQUID (Superconducting Quantum Interference Device, basically a sliced version of the cylinder we are considering).
  \item These proposed devices might be bulkier than an average transistor fabricated and might be trickier to fabricate than regular CMOS as this does not favour a planar geometry. Also they are not scalable as their CMOS counterparts.
  \item Noise will definitely favour they proposed class of circuits as they are operating at 2 orders of temperature lower than regular CMOS. Thus its not fair to compare thermal noise performance.
  \item However these kind of circuits are almost immune to the concept of offset and mismatch.
\end{itemize}
In following secitons, we will concentrate more on the amplification and transient performances. Noise will however be dealt with in Appendix-A.

\begin{figure*}[!t]
\centering
\setlength\fboxsep{0pt}
\setlength\fboxrule{0.25pt}
\fbox{\includegraphics[width=\textwidth,keepaspectratio=true]{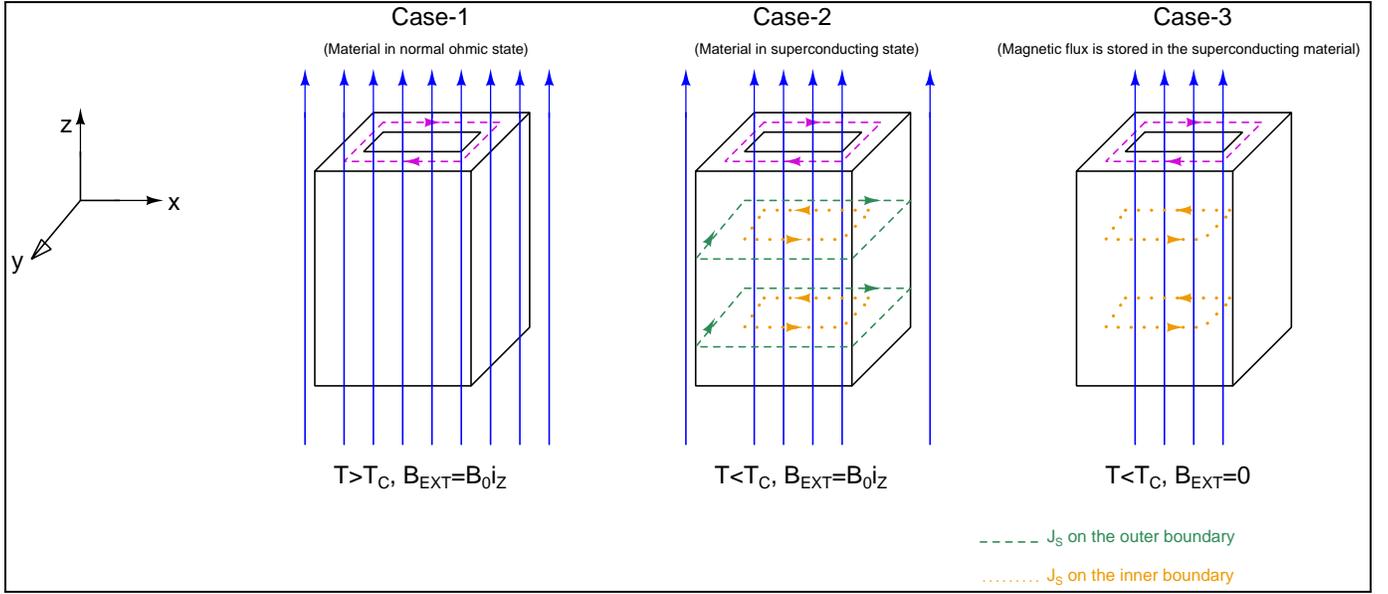}}
\caption{Flux trapping in a superconductor based on Meissner effect and classical Electrodynamics.}
\label{flux_trap}
\end{figure*}

\section{Working of the Proposed Circuits and Devices}
We will begin by borrowing the concept of magnetic diffusion in a material present in magnetic field. From electromagnetodynamics and applying Gauss's law of magnetic field divergence
\begin{equation}
\nabla \times \nabla \times \vec{B} = -\nabla^2 \vec{B} = \mu_0 \nabla \times \vec{J}
\end{equation}
Applying Ohm's Law in a conducting medium stationary with respect to a magnetic field
\begin{equation}
\vec{E} = \sigma \vec{J}
\end{equation}
rearranging which gives
\begin{equation}
\bigg( \mu_0\sigma_0 \frac{\partial}{\partial t}  -  \nabla^2 \bigg) \vec{B} = 0
\end{equation}
Now referring to Fig. \ref{flux_trap} suppose we have a conductor which extends to infinity in \^y and \^z directions but has finite width in \^x direction. It is placed in a region in space with uniform $\vec{B}$ directed in \^z given by
\begin{equation}
\vec{B}_{ext} = Re\bigg\{ \vec{B}_{0z} e^{j\omega t} \bigg\}
\end{equation}
Solving for the magnetic field inside the conductor and the circulating surface current from the above two equations are
\begin{equation}
\vec{B}_{int} = Re\bigg\{ \vec{B}_{0z} \frac{cosh (1+j)\sqrt{\frac{\omega \mu \sigma}{2}}x}{cosh (1+j)\sqrt{\frac{\omega \mu \sigma}{2}}d} e^{j\omega t} \bigg\}
\end{equation}
where 2d is the finite thickness of the slab in \^x, $\omega \: \mu \: \sigma$ are properties of the material of the slab and
\begin{equation}
\vec{J}_{circ} = Re\bigg\{ \sqrt{\frac{\omega \sigma}{2\mu}} \vec{B}_{0z}(1+j) \frac{sinh (1+j)\sqrt{\frac{\omega \mu \sigma}{2}}x}{sinh (1+j)\sqrt{\frac{\omega \mu \sigma}{2}}d} e^{j\omega t} \bigg\}
\end{equation}

\par

\begin{figure*}[h]
\centering
\setlength\fboxsep{0pt}
\setlength\fboxrule{0.25pt}
\fbox{\includegraphics[width=\textwidth,keepaspectratio=true]{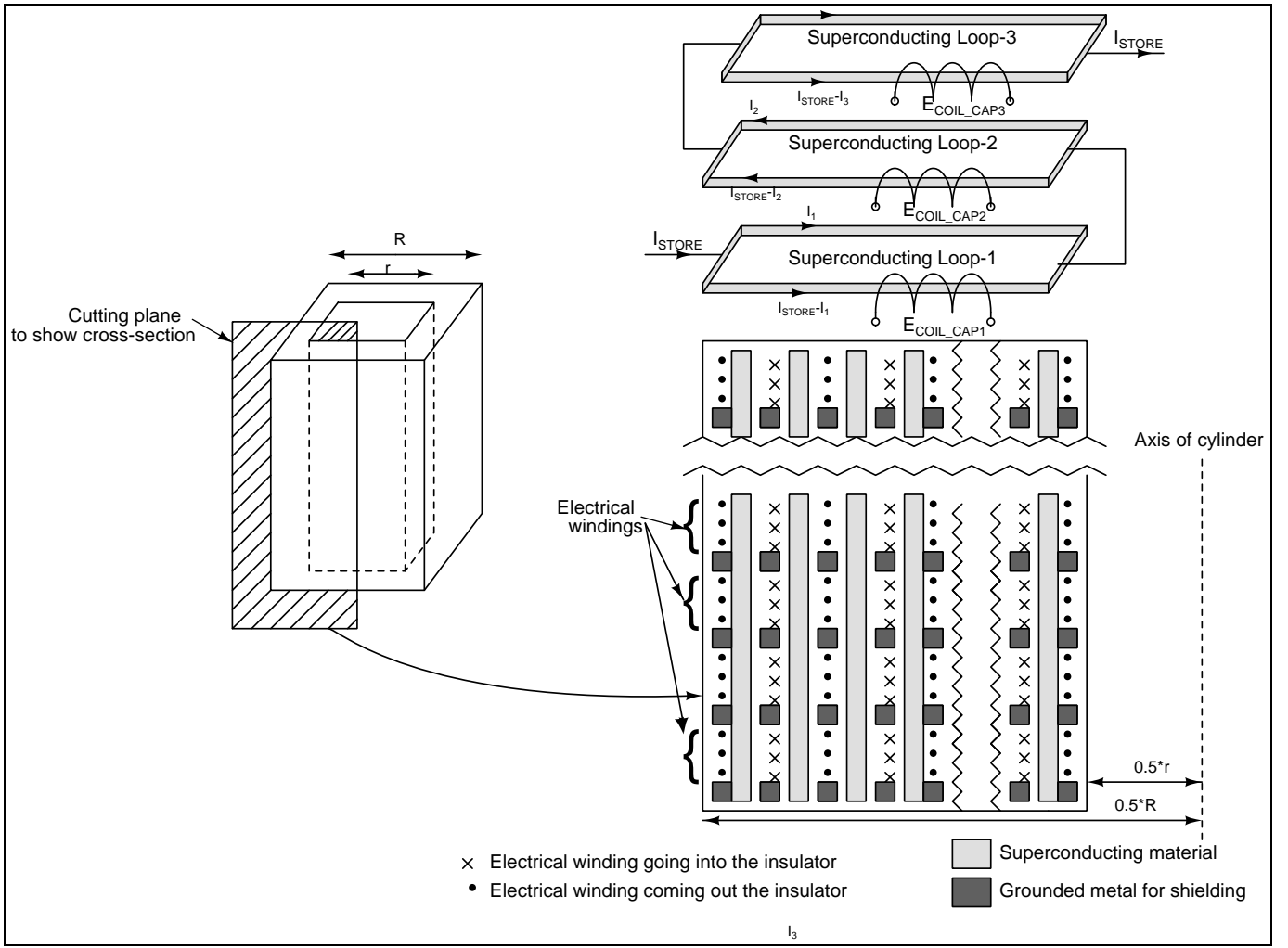}}
\caption{Construction of the magnetic flux trapping and amplifying+integrating in a superconductor.}
\label{amp_integ}
\end{figure*}

However, for a superconducting medium placed in a magnetic filed lower than $\vec{H_C}$ for type-\rom{1} or $\vec{H_{C2}}$ for type-\rom{2} we will have a slightly different set of equations. First of all we get a modified Ohm's Law for superconductors proposed by the London brothers and given by
\begin{equation}
\vec{E} = \frac{\partial}{\partial t}(\Lambda \vec{J})
\end{equation}
The magnetic diffusion equation is obtained by proceeding from
\begin{equation}
\frac{\partial}{\partial t} \bigg( \nabla \times \nabla \times \vec{B} \bigg) = -\frac{\partial}{\partial t} \bigg( \nabla^2 \vec{B} \bigg) = \mu_0 \nabla \times \frac{\partial \vec{J}}{\partial t}
\end{equation}
rearranging which gives
\begin{equation}
\bigg( \frac{1}{\Lambda}  -  \frac{\nabla^2}{\mu_0} \bigg) \frac{\partial}{\partial t}\vec{B} = 0
\end{equation}
Again assuming the same orientation of superconductor and magnetic field as before, we arrive at
\begin{equation}
\vec{B}_{int} = Re\bigg\{ \vec{B}_{0z} \frac{cosh\:x\sqrt{\frac{\mu_0}{\lambda}}}{cosh\:d\sqrt{\frac{\mu_0}{\lambda}}} e^{j\omega t} \bigg\}
\end{equation}
where 2d is the finite thickness of the slab in \^x, $\omega \: \mu \: \sigma$ are properties of the superconducting material of the slab and
\begin{equation}
\vec{J}_{circ} = Re\bigg\{ \frac{1}{\sqrt{\mu \lambda}} \vec{B}_{0z} \frac{sinh\:x\sqrt{\frac{\mu_0}{\lambda}}}{sinh\:d\sqrt{\frac{\mu_0}{\lambda}}} e^{j\omega t} \bigg\}
\end{equation}
If by design we are in such a state where the current circulating inside a superconducting block is composed of both normal carriers$\rightarrow$electrons and supercarriers$\rightarrow$superelectrons, then the equation describing the complete distribution over time and length scales is
\begin{equation}
\Bigg[ 1 - \lambda^2\nabla^2 + \bigg( \mu\sigma\lambda^2 - \tau_s\lambda^2\nabla^2 + \tau_s \bigg)\frac{\partial}{\partial t} \Bigg] \vec{B} = 0
\end{equation}
where $\tau_s$ is the scattering time-constant of the superconductor. However such situations will only arise when transitioning from superconducting pahse to normal phase or vice-versa. Now we will look at how this stored magnetic is magnified and integrated. In the Fig. \ref{amp_integ} a cylinder is shown contructed from a type-\rom{1} superconductor.\par
After we reach the trapped flux state of the rightmost case in Fig. \ref{flux_trap}, the supercurrents will circulate all over the height of the cylinder. Instead if we selectively make some regions of the superconducting annulus ohmic and only a part as superconducting, this circulating current will be limited to only those portions which are in superconducting phase. This can be achieved by sending currents into the coils shown by \ding{53} and \ding{108} which generate an axial magnetic field through the superconductor strong enough to break the Cooper-pairs and convert the superconductor locally under the coil into normal ohmic metal. Thus the trapped magnetic flux will be maintained by circulating supercurrents flowing in only certain section of the superconductor. The supercurrent can be given by the following approximate solenoidal relation
\begin{equation}
\vec{B}_{trapped} = \mu n I_{S}
\end{equation}
which is not valid at the upper and lower edges of the cylinder due to fringing. Here n is the effective turns per unit length and $I_{S}$ is the total supercurrent flowing in the part which is still superconducting. Now let us focus on a small section of Fig. \ref{amp_integ} consisting of two windings separated by a grounded metal shielding with the superconducting material enclosed within. Let us also consider that only the top coil is carrying a current such that the portion enclosed by it is not superconducting whereas the lower coil is not carrying any current and thus the portion enclosed by it is still superconducting. After establishing such a condition, if we go through the steps shown in Fig. \ref{flux_trap}, the circulating supercurrent will be flowing only in the lower portion which is still superconducting and not the upper part. Then if we stop the current in the upper coil, the material enclosed will return to the normal superconducting state and the supercurrent will spread over twice the original area to satify the Maxwell-Faraday equation
\begin{equation}
\mathlarger{\sum_{1}^{S}} \oint_C \vec{E} \,dl = \mathlarger{\sum_{1}^{S}} \Bigg[ - \frac{\partial}{\partial t} \int_S \vec{B} \,dS \Bigg]
\end{equation}
with the difference being, here we consider miltiple closed contours in the void enclosed by the total cylinder defining a surface through which the magnetic field needs to be constant. Here S is the number of contours we are considering in the given space between end to end of two coils; the answer remains the same independent of the number of such contours chosen. Thus evidently we see that after turning on both the coils, the supercurrent density ($\vec{J}_S$) becomes almost half but the supercurrent ($\vec{I}_S$) remains the same. Next if we turn the current through the lower coil only thus making the lower portion of the enclosed material normal, the supercurrent will flow only in the upper portion where the material is still superconducting. By doing this whole sequence of turning on/off sequentially each coils we are able to bring such "discs" of superconducting currents into the same horizontal plane. Because the current flowing in each strand of superconductor is bounded on either sides and not have dissipated, we now have an amplification of the trapped magnetic field by a factor $\bf{N}_{\bf{SC}}^{\bf{*}}$ where $\bf{N}_{\bf{SC}}^{\bf{*}}$ is the number of stacks of coil that can be manufactured into the same cylinder. Based on our working knowledge of the device developed so far we can readily draw the following conclusions on the amplification factor $\bf{N}_{\bf{SC}}^{\bf{*}}$ -: 
\begin{itemize}
  \item We can have an amplification factor that is totally independent of fabrication imperfections like process, mismatch, gradients, etc.
  \item We will see in later sections that this factor depends on only quantum mechanical and thermodynamic constants thus making it robust.
  \item In classical CMOS circuits we could achieve higher gains by cascading amplifier stages and increasing headroom. However this posed a stability issue as most of the times such structure were operated in closed loop. With our proposed structure we are not bound by stability and can theoretically cascade such structures to resemble a distributed amplifier with high enough gain.
  \item Also this amplification factor does not have any low frequency problems like drift which plague classical CMOS circuits.
\end{itemize}
If this device is to be used as an amplifier, we can do so easily by placing a SQUID on top of the cylinder but isolated from it electromagnetically. So once one set of amplificaiton sequence is done, the resultant magnetic field is stored in the SQUID. The cylinder is reset and next amplification sequence begins. Post that the resultant magnetic field is again added to the SQUID and the overall device performs like an integrator with gain. For more detailed description of how a SQUID functions, one can refer to the books mentioned in reference section. Thus we have formulated a replacement of those building blocks marked in saffron in Fig. \ref{classical_DSM}. The feedback digital-to-analog converter block shown in the same figure can also implemented by passing the reference current through a solenoid as can be done with the input current too. The only blocks that are now left are the comparator and the feedback-DAC which will be taken up in the last section.\par
Shown below is a similar construction of the cylinder but with a type-\rom{2} superconductor which shows the formation of vortices for fields higher than first critical field.
\begin{figure}[!ht]
\includegraphics[scale=0.38]{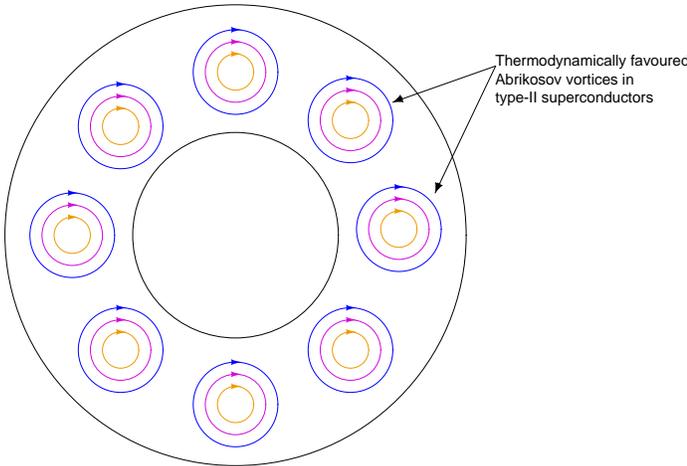}
\caption{Presence of Abrikosov vortices in type-\rom{2} superconductor for magnetic fields higher than H\textsubscript{C1}$<$H\textsubscript{EXT}$<$H\textsubscript{C2}.}
\label{amp_integ_2}
\end{figure}
Here too the same mathematics and arguments hold as would in the case of a type-\rom{1} superconductor discussed previously in this section.\par
There remains one final thing in this section, applicable to both type-\rom{1} and type-\rom{2} superconductors, which expresses the magnetic field trapped inside a superconductor is quantized. This is not true for any normal fields generated by a current carrying wire or a solenoid; those can assume any particular value but not the trapped field inside a superconductor. This concept of quantization is foreign to the classical domain and we will need to invoke quantum electromagnetics to explain it. Consider the following Schr\"{o}dinger wave equation in the simplest form
\begin{equation}
i\hbar \frac{\partial \psi}{\partial t} = - \frac{\hbar^2}{2m}\nabla^2\psi + V(x,y,z,t)\psi
\end{equation}
We define a probability density of the aforementioned wavefunction $\psi$ as
\begin{equation}
\frac{\partial \rho_S}{\partial t} = - \nabla J_{\rho}
\end{equation}
where $\rho_S$ is the given by $\left|\psi^2\right|$  as defined by Born to be the probability of finding the particle at a given location and a given time, $J_{\rho}$ is the associated probability current of the particle/s described by $\psi$. Thus $J_{\rho}$ is given by
\begin{equation}
J_{\rho} = - \frac{\hbar}{2im}( \psi^*\nabla\psi - \psi\nabla\psi^* ) = Re \bigg\{ \psi^{*} \frac{\hbar}{im} \nabla \psi \bigg\}
\end{equation}
In our specific case involving the motion of charged particle in electric and magnetic fields, we need to find a quantum-mechanical version of the classical Lorentz's Force law which can be readily obtained from textbooks dealing in such given by the following classical and quantum-mechanical versions
\begin{equation}
m\frac{d\vec{v}}{dt} = q \{ \vec{E} + (\vec{v}\times\vec{B}) \}
\end{equation}
\begin{equation}
i\hbar \frac{\partial \psi}{\partial t} = - \frac{1}{2m}\bigg[ \frac{\hbar}{i}\nabla -qA(x,y,z,t)\bigg]^2\psi + q\phi(x,y,z,t)\psi
\end{equation}
For such a wavefunction $\psi$ the probability current is given by
\begin{equation}
J_{\rho} = Re \bigg\{ \psi^{*} \bigg[\frac{\hbar}{im}\nabla - \frac{qA(x,y,z,t)}{m}\bigg]\psi \bigg\}
\end{equation}
If we define the constituents of the supercurrent using an overall macroscopic many-body wavefunction instead of individual wavefunctions for wach superelectons
\begin{equation}
\Psi(x,y,z,t) = \sqrt{n^*(x,y,z,t)}e^{i\theta(x,y,z,t)}
\end{equation}
where $n^*(x,y,z,t)$ is the local density of superelectrons. Using this expression for $\Psi$ in the previous equation, we get
\begin{equation}
J_S^* = q^*n^*(x,y,z,t)\bigg[ \frac{\hbar}{m^*}\nabla\theta(x,y,z,t) - \frac{q^*}{m^*}A(x,y,z,t) \bigg]
\end{equation}
where the term in the brackets is the effective superelectrons velocity. After rearranging and some algebraic manipulations we get
\begin{equation}
\Lambda J_S^* = - \bigg[A(x,y,z,t) - \frac{\hbar}{q^*}\nabla\theta(x,y,z,t) \bigg]
\end{equation}
Remembering the definition of magnetic vector potential and electric potential
\begin{equation}
\vec{B} = \nabla\times\vec{A} , \; \; \; \; \;  \vec{E} = -\nabla\phi - \frac{\partial\vec{A}}{\partial t}
\end{equation}
Integrating the prior equation we get
\begin{equation}
\oint_C(\Lambda\vec{J}_S^*).\vec{dl} + \int_S \vec{B} \vec{dS} = \frac{\hbar}{q^*}\oint_C\nabla\theta.\vec{dl}
\end{equation}
In general integration of a spatial derivative over a countour is simply given by
\begin{equation}
\int_{r_a}^{r_b} \nabla\theta.\vec{dl} = \theta(r_b,t) - \theta(r_a,t)
\end{equation}
if $r_a$$\rightarrow$$r_b$ such that a closed contour is traced the the integral evaluates to zero. However if the function $\theta$ is composed of a particular and a general solution given by $\theta(x,y,z,t) = \theta_P(x,y,z,t) + 2\pi n$ where n is an integer, then the solution to the above integral is just $2\pi n$. Thus we see that the RHS of the integration gives $\frac{hn}{q_S^*}$ which shows that the magnetic field flux trapped inside the hollow of our superconducting cylinder is indeed quantized by $\phi_0=\frac{h}{q_S^*}$ = 2.0706 femto-Tesla$\cdot$$m^2$.\par
To get an idea about the speed and performance comparison of the proposed device and its classical counterpart let us assume that we have a superconducting cylinder as shown in Fig. \ref{amp_integ} wherein by construction we have $N_{AMP}$ rings. Also let us assume that the magentic field trapped inside the superconducting cylinder needs a finite settling time based on the mobility of the Cooper-pairs among other things $\rightarrow$ $\tau_{cooper}$. In any typical sampled system the time taken to settle to $\frac{1}{2}$LSB accuracy for a ``N-bits'' system is
\begin{equation}
t_{settle} = \tau_{RC}(N+1)\ln(2)
\end{equation}
where R and C are the total path resistance and sampling capacitor of the network. So for a 22-bit linearity system we would require about 16 time constants. Assuming the engaging and disengaging of each of the E-coils takes $\tau_{E-coil}$, we would need a total time for the settling of the integrator around
\begin{equation}
\mathbf{t_{settle}^*} = \bigg[16\tau_{cooper}\ln(2)\bigg] + \bigg[(2N_{AMP}+1)\times\tau_{E-coil}\bigg]
\end{equation}
For the purpose of illustration, we note that the $2^{nd}$-term in the above equation dominates. Assuming the digital standard-logic gates are in lower technology nodes, we can roughly conclude the settling-time to be of the order of 20$\,$nano-seconds. If the same was to be achieved using a classical CMOS opamp based discrete time integrator with a transfer function
\begin{equation}
A_V = \frac{C_i}{C_f}\bigg(\frac{z^{-1}}{1-z^{-1}}\bigg)
\end{equation}
we would require to burn about 1.12mA in just the input differential-pair to meet 20ns settling time across corner and temperature assuming C=2pF and feedback factor $\beta=\frac{1}{10}$. Whereas in the proposed device it would primarily be the current required by the digital control gates driving the E-coils. Overall current benefits would be compared with existing ADC architectures in table-\rom{1}.\par
\begin{figure}
\centering
\includegraphics[width=\columnwidth,height=4.5in]{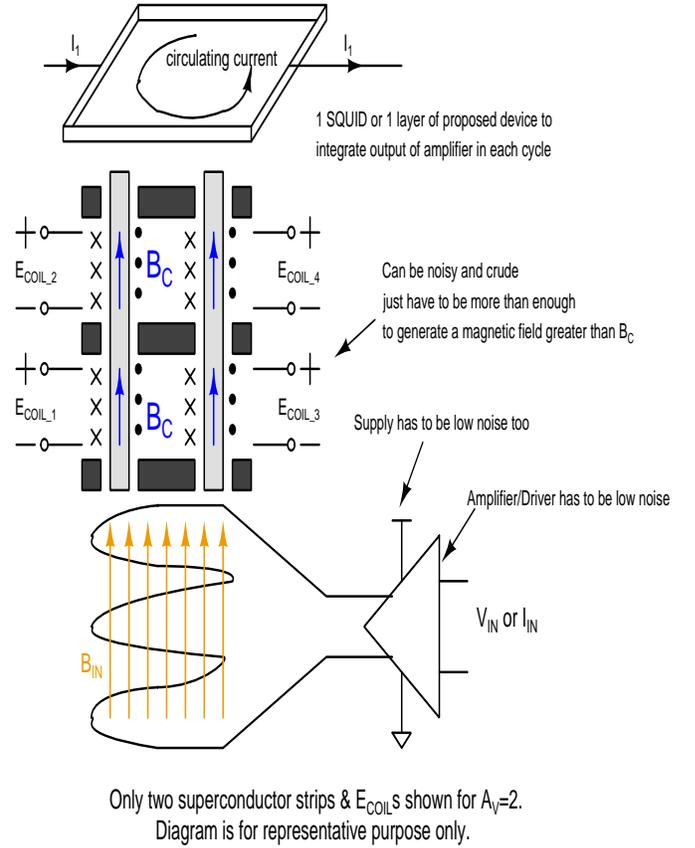}
\caption{A typical arrangement showing the amplifier with an input coil.}
\label{device_func}
\end{figure}

For those conversant in the art of electronic circuit design, the arangement shown in Fig. \ref{device_func}. The operation can be easily understood by following the outlined steps.
\begin{itemize}
  \item We begin by turning $E_{COIL}$[1:4] on and the input solenoid also on.
  \item Thing to note here is that $B_{IN}$ has to be lower than the absolute magnetic field of the superconductor material.
  \item Next, only $E_{COIL}$[1] and $E_{COIL}$[4] are turned off, transitioning the respective segments into superconducting regions. Rest remain in ohmic regions.
  \item Now the input solenoid is turned off making $B_{IN}$=0.
  \item The magnetic field information of $B_{IN}$ is stored by the circulating supercurrent inside the superconducting segments 1 and 4.
  \item Next keeping $E_{COIL}$[3] on, we turn off $E_{COIL}$[2] thus allowing the current a larger cross-section to flow or circulate.
  \item Finally we turn on $E_{COIL}$[1] restricting the two concentric superconductor currents in segments 2 and 4. Thus we can magnify $B_{IN}$ by a factor of two.
\end{itemize}
\par
Another probable method of implementing a gain stage is shown below in Fig. \ref{two_coils}
\begin{figure}
\centering
\includegraphics[width=\columnwidth]{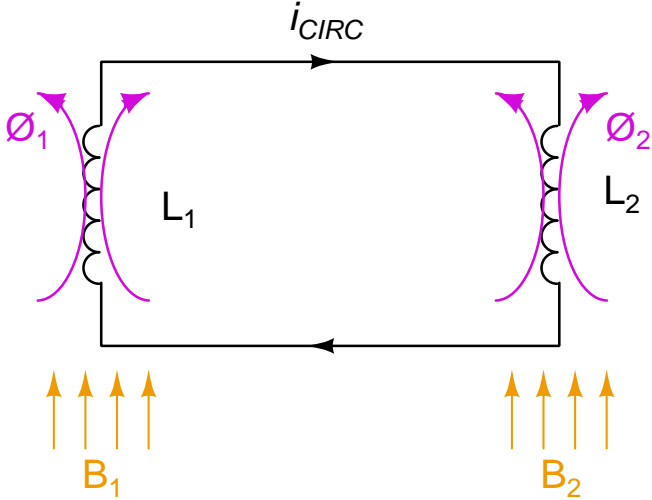}
\caption{Arrangement showing the storage of magnetic flux in connected superconducting coils.}
\label{two_coils}
\end{figure}
The time derivative of the total flux linked by the coils will be constant as we will show below. Consider we begin with the loops being in normal ohmic state which can be easily done using the E-coil arrangement shown in the prior amplifiers. This makes sure that the flux incident on the loops do penetrate through. After the flux threading the loops settle with a time constant, the loops are again brought back to superconducting state using the same E-coil action. The total circulating current $i_{CIRC}$ is given by
\begin{equation}
\frac{d}{dt}(\lambda_1 + \lambda_2) = -i_{CIRC} \times R_{LOOP}
\end{equation}
But due to the perfect conduction properties, the electric field inside the superconductor is zero ($\vec{E}=0$) and thus $R_{LOOP}$ is also zero. This means the total incident flux at the time instant when the loop became superconducting is now ``sampled'' or ``frozen''. This idea is now extended to multiple such loops or woundings shown in Fig. \ref{multiple_coils}. Initially all the coils are exposed to the external magnetic field $\vec{B}_{IN}$ which leads us to the following equation
\begin{equation}
N\lambda_0 + (N-1)i + L_{AMP}i_{AMP-I} = k
\end{equation}
where $\lambda_0$ is $\vec{B}_{IN}\times A_{loop}$ and k is a constant. Next we reduce the magnetic fields to the loops from $\vec{B}_{IN}$ to $\epsilon\vec{B}_{IN}$. The loops or inductors being linear, the current in them would also scale down by the same amount i.e. from $i$ to $\epsilon i$. Hence we would have
\begin{equation}
N\epsilon\lambda_0 + (N-1)\epsilon i + L_{AMP}i_{AMP-F} = k
\end{equation}
From the above two equations we get
\begin{equation}
\Delta\lambda_{AMP} = \lambda_{AMP-F} - \lambda_{AMP-I} = N(1-\epsilon)\lambda_0 + (N-1)(1-\epsilon)Li
\end{equation}
The first term gives us the amplification of input flux while the second term can be easily calibrated out using digital post-processing. Typically the fields would be turned off in an application which would lead to $\epsilon=0$. In other words when the flux inside the inductors [$L_1 \rightarrow L_N$] are simultaneously changed, the flux through $L_{amplify}$ changes in such a fashion that the initial flux given by $B_{IN} \times A_{loop}$ remains constant. In other words, if the net flux through [$L_1 \rightarrow L_N$] are made zero, we can get a amplified field through the $L_{amplify}$ inductor. The inductors can be easily laid out in a fashion similar to the ones shown in the previous amplifiers.
\begin{figure}
\centering
\includegraphics[width=\columnwidth]{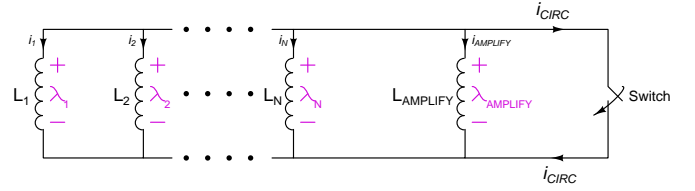}
\caption{Arrangement showing multiple coils whose stored flux can be made to move to the amplifying coil creating higher magnetic flux.}
\label{multiple_coils}
\end{figure}

\section{Transient control of Superconductivity through controlling coils}
In this section we will look at the transient behaviour of the device when the current carrying coils are sequentially turned on or off to control superconductivity. From here onwards we will refer to these coils as E-coils. Shown below are the strengths of external magnetic fields beyond which a type-\rom{1} superconductor becomes normal ($H_{C1}$ or $H_C$) and beyond which a type-\rom{2} superconductor becomes normal ($H_{C2}$). The E-coils when energized locally make the superconductor back to normal by this action of higher magnetic field passing through them. Following figure shows the temperature dependence of these critical field limits on temperature.
\begin{figure}[!ht]
\includegraphics[scale=0.5]{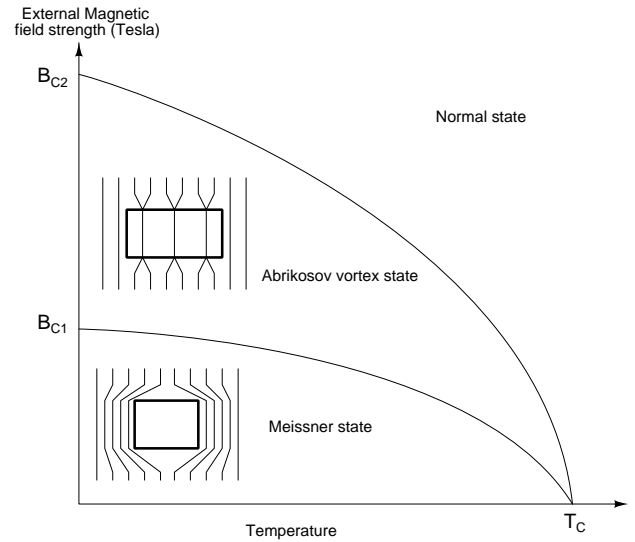}
\caption{Variation of critical fields in type-\rom{1} and type-\rom{2} superconductors with temperature.}
\label{sc_normal_field}
\end{figure}
So our current problem becomes that of modelling the ``normal-superconducting" junction. To do so we turn to the Hartree-Fock method of approximation of the many-body wavefunction by a permanent in case of bosons (as our current case). To cope with the problems which arise when the superconducting order parameter varies spatially like in the case of a junction or a vortex, physicists have used the Bogoliubov-de Gennes equations which extend the Hartree-Fock equations to include the rairing potential $\Delta$($\vec{x}$) as well as the ordinary magnetic and non-magnetic potentials. However, before we begin, we would like to briefly describe the carrier concentration on either side of the junction. This is described using the following figure.
\begin{figure}[!ht]
\includegraphics[scale=0.75]{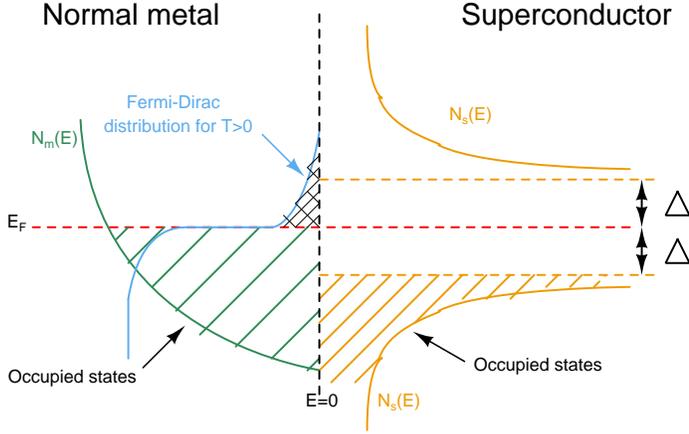}
\caption{Energy-position diagram showing the desity of states and carrier concentration and band occupation on both sides of the junction for temperatures greater than 0K but less than $T_C$.}
\label{band_structure}
\end{figure}
Here the electrons can tunnel from the normal side to the superconducting side to available energy-states in two ways. First is the normal mode where the thermally excited electons move to the density of states which are $\Delta$ above the Fermi-energy. Notice the bandgap-kind of structure characteristic of a semiconductor; difference being the denisty of states N(E) does not diverge near the bandgap, instead is follows a parabolic increase with energy. These electrons having much higher energy than the normal Cooper-pairs do not take part in the circulating supercurrent and thus do not affect the magnitude of the trapped magnetic field. Second and more important mode are when the lower energy electrons below $E_F$-$\Delta$ tunnel to the superconducting side. An electron propagating in such a way experiences a special type of scattering called Andreev reflection wherein a hole is reflected into the metal with opposite momentum and a Cooper-pair is generated in the superconductor side.\par
The Bogoliubov-de Gennes equations for the motion of particles in ``normal-superconducting" interface become:
\begin{equation}
\begin{split}
-\frac{\hbar^2}{2m}\bigg(\nabla-\frac{ie}{\hbar c}\vec{A}\bigg)^2\!\vec{u} \;+\; [\vec{U_{ex}} - E_F]\vec{u} \;+\; \vec{\Delta}\vec{v} = \epsilon\vec{u}
\end{split}
\end{equation}
\begin{equation}
\begin{split}
\frac{\hbar^2}{2m}\bigg(\nabla+\frac{ie}{\hbar c}\vec{A}\bigg)^2\!\vec{v} \;-\; [\vec{U_{ex}} - E_F]\vec{v} \;+\;  \vec{\Delta^*}\vec{u} = \epsilon\vec{v}
\end{split}
\end{equation}
Considering the system to be uniform in $\hat{y}$ and $\hat{z}$ directions we can define u and v as
\begin{equation}
\begin{bmatrix} \vec{u} \\ \vec{v} \end{bmatrix} = e^{ik_y\hat{y} + ik_z\hat{z}} \begin{bmatrix}u(x) \\ v(x) \end{bmatrix}
\end{equation}
Inserting them into the BdG equations, assuming anisotropic $\Delta$ with no surface imperfections or barriers, we get
\begin{equation}
\bigg[-\frac{\hbar^2}{2m}\frac{d^2}{dx^2} - E_x\bigg]u_x + \Delta v_x = \epsilon u_x
\end{equation}
\begin{equation}
-\bigg[-\frac{\hbar^2}{2m}\frac{d^2}{dx^2} - E_x\bigg]v_x + \Delta^* u_x = \epsilon v_x
\end{equation}
where 
\begin{equation}
E_x = E_F - \frac{\hbar^2(k_y^2+k_x^2)}{2m}
\end{equation}
and the momentum vector traces out a sphere in momentum space.
At the boundary the wavefunctions $\textit{\textbf{u}}$, $\textit{\textbf{v}}$ and their derivatives $\textit{\textbf{u'}}$, $\textit{\textbf{v'}}$ must be continuous given by
\begin{equation}
\begin{bmatrix} u_x(0) \\ v_x(0) \end{bmatrix}_{normal} = \begin{bmatrix}u_x(0) \\ v_x(0) \end{bmatrix}_{superconducter}
\end{equation}
\begin{equation}
\begin{bmatrix} u'_x(0) \\ v'_x(0) \end{bmatrix}_{normal} = \begin{bmatrix}u'_x(0) \\ v'_x(0) \end{bmatrix}_{superconducter}
\end{equation}
For the sake of keeping it concise we will state the results from the above equations directly for the four use cases of when an electron or a hole is incident on the interface from the normal or the superconducting side. For a particle excitation with energy $\epsilon>\Delta$:
\begin{equation}
\begin{bmatrix} u_x \\ v_x \end{bmatrix}_{normal} = e^{ik_x^{N_x}}\begin{bmatrix}1\\0\end{bmatrix} + \frac{U}{V}e^{ik_x^{-N_x}}\begin{bmatrix}0\\1\end{bmatrix}
\end{equation}
\begin{equation}
\begin{bmatrix} u_x \\ v_x \end{bmatrix}_{superconductor} = \frac{1}{U}e^{ik_s^{N_s}}\begin{bmatrix}U\\V\end{bmatrix}
\end{equation}
here U and V are called the coherence factors in a uniform superconductor given by
\begin{equation}
\begin{split}
U = \frac{1}{\sqrt{2}} \bigg(1+\frac{\sqrt{\epsilon^2 - \Delta^2}}{\epsilon} \bigg)^\frac{1}{2} \\
V = \frac{1}{\sqrt{2}} \bigg(1-\frac{\sqrt{\epsilon^2 - \Delta^2}}{\epsilon} \bigg)^\frac{1}{2}
\end{split}
\end{equation}
For the case when the particle excitation with energy $\epsilon<\Delta$ is incident on the interface, the same equations for $u_x$ and $v_x$ are still valid but U and V are modified:
\begin{equation}
\begin{split}
U = \frac{1}{\sqrt{2}} \bigg(1+i\frac{\sqrt{\Delta^2 - \epsilon^2}}{\epsilon} \bigg)^\frac{1}{2} \\
V = \frac{1}{\sqrt{2}} \bigg(1-i\frac{\sqrt{\Delta^2 - \epsilon^2}}{\epsilon} \bigg)^\frac{1}{2}
\end{split}
\end{equation}
From the above equations we come to the following conclusions for the four cases of incidence -:
\begin{itemize}
  \item If the incident particle is an electron from the metal side then its anti-particle (hole) is Andreev reflected back into the metal while a similar particle (an electron) is generated in the superconductor and moves in the transmission direction. There are no specular refelction components.
  \item If the incident particle is a hole from the metal side then its anti-particle (electron) is Andreev reflected back into the metal while a similar particle (a hole) is generated in the superconductor and moves in the transmission direction. Here too there are no specular refelction components.
  \item If the incident particle is an electron with energy $\epsilon<\Delta$ from the metal side then its anti-particle (hole) is Andreev reflected back into the metal while a Cooper-pair is generated in the superconductor and moves in the transmission direction with no specular refelction components.
\end{itemize}
The contribution of this effect to noise in the circulating supercurrent inside the superconductor will be taken up in Appendix-A. We will now end this section with the derivation of currents originating from the above tunnelling process.
\par
Considering we have particle and hole excitations given by something similar to eqn(31) and inserting them into the first BdG equation would give
\begin{equation}
\begin{split}
-\frac{\hbar^2k_F}{2m}\bigg[ 2i\vec{k}\nabla U(x) + \frac{2e}{\hbar c}\vec{k}\vec{A}U(x) - \frac{2ie}{\hbar k_Fc}\vec{A}\nabla U(x) \\ -\frac{e^2}{k_F\hbar^2 c^2}\vec{A}^2U(x) + \frac{1}{k_F}\nabla^2U(x) \bigg] = \epsilon U(x)
\end{split}
\end{equation}
Using the approximations $\frac{1}{k_K}\approx a_0$ (the de Broglie wavelength associated with inter-atomic distance), $\nabla U(x)\approx U(x)/\xi$, $\nabla^2U(x)\approx U(x)/\xi^2$, $e\vec{A}/(\hbar c)\approx 1/\xi$ and $\vec(A)\approx \frac{\hbar c}{e\xi}$ we can neglect the last three terms in the bracket obtaining the Andreev equations:
\begin{equation}
-i\hbar v_F \bigg( \nabla - \frac{ie}{\hbar c}\vec{A} \bigg)U + \Delta V = \epsilon U
\end{equation}
\begin{equation}
i\hbar v_F \bigg( \nabla + \frac{ie}{\hbar c}\vec{A} \bigg)V + \Delta^* U = \epsilon V
\end{equation}
The solution to this in the normal region where $\Delta$ is zero would give
\begin{equation}
\begin{bmatrix} u_x \\ v_x \end{bmatrix}_{normal} = \begin{bmatrix} a_0e^{i(\frac{\epsilon}{\hbar v_F}+\frac{e\vec{A}}{\hbar c})x} \\ a_1e^{-i(\frac{\epsilon}{\hbar v_F}+\frac{e\vec{A}}{\hbar c})x} \end{bmatrix}
\end{equation}
where $a_0$ and $a_1$ are constants of integration. Now the quantum supercurrent flowing through the normal-superconductor junction would have a supercurrent density given by
\begin{equation}
\mathbf{J_S^*} = \frac{e}{m}\mathlarger{\sum_{n}} \bigg[f(\epsilon_n)(u_n^*\hat{p}u_n + u_n\hat{p}^{\dagger}u_n^*)  +  
(1-f(\epsilon_n))(v_n\hat{p}v_n^* + v_n^*\hat{p}^{\dagger}v_n) \bigg]
\end{equation}
summed over n quantum states, $f(\epsilon)$ is the Fermi-Dirac distribution and $\hat{p}$ is the canonical momentum operator. Finally, skipping the rigorous mathematics and jumping directly to the result for long junctions gives
\begin{equation}
\begin{split}
\mathbf{I_{SN}^*} & = \frac{2eSv_Fk_F^2}{\pi^2d} e^{-d/\xi_N}sin\phi \\
								    & = \frac{4\hbar N(0)v_F^2eS}{d} e^{-d/\xi_N}sin\phi \\
								    & = \frac{16\hbar v_F}{2edR_{SH}} e^{-d/\xi_N}sin\phi \\
\end{split}
\end{equation}
where $R_{SH}$ is the resistance coming from the normal-superconductor contact, $\xi_N = \frac{\hbar v_F}{2\pi k_BT}$ gives the distance over which the superelectrons' current decays into the normal metal.

\section{Bose-Einstein statistics in presence of external Electromagnetic fields}
In this section we will relook at the seminal Bose-Einstein statistics for bosons in an interacting electromagnetic field which in our case are the Cooper pairs.

$<<$This is still a work in progress and will be added when the underlying mathematics is on a firmer footing.$>>$

\section{Comparator and Feedback-DAC Design}
The comparator can be a superconducting ring made of the same material as that used for the amplifier-integrators in the loop-filter. A planar 2-D surface encompassing the total cross-sectional area of the amplifying+integrating cylinders can be designed and placed at the same axis as the cylinders themselves. 
\begin{figure}[H]
\includegraphics[scale=0.75]{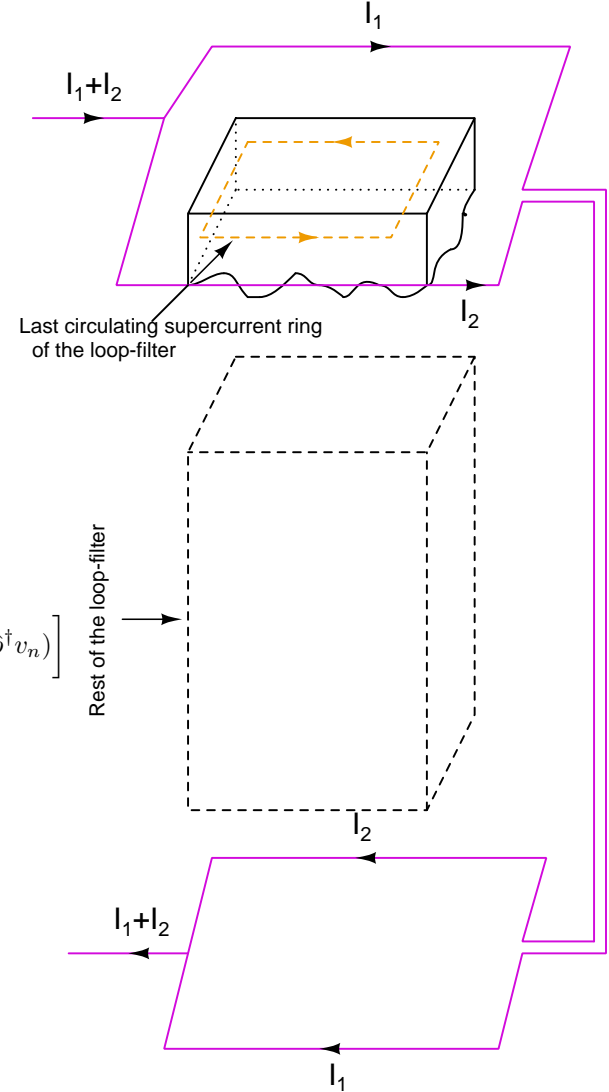}
\caption{Construction of the comparator.}
\label{Comparator+DAC}
\end{figure}
While the loop-filter is integrating, the comparator superconducting loop can be made to be in a normal mode either by the action of suitably placed E-coils or by pumping a current more than the critical current $I_C$ such that superconductivity is lost. Later on after the loop-filter is finished with the amplification of the current cycle, the comparator loops can be allowed to regain superconductivity. It is at this time that the final output of the loop-filter is stored in the comparator loops by the differential circulating currents depending upon the magnetic field. From Biot–Savart law we have the magnetic field at the center of the loop as
\begin{equation}
\frac{\vert I_1 - I_2 \vert}{2} = \frac{\pi L \vec{B}_{LF}}{2\sqrt2\mu_0}
\end{equation}
where $L$ denotes the lenght of one side of the comparator loops and $\vec{B}_{LF}$ is the output magnetic field from the loop-filter. The difference in current in the two arms is to maintain the resultant magnetic field.
\begin{figure}
\centering
\includegraphics[scale=0.86]{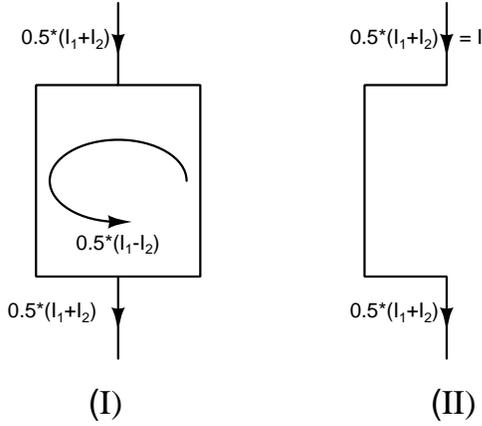}
\caption{Resultant magnetic field inside the comparator loop during normal operation (\rom{1}) and when saturated (\rom{2}).}
\end{figure}
From this relation we can easily conclude a few facts
\begin{itemize}
  \item The maximum magnetic-field that the comparator loop can detect without overload is when it is so strong that the entire bias current for the comparator loop is diverted to one side.
  \item The minimum magnetic field that can be detected reliably would be the single quanta of magnetic flux given by $B_{LSB}=\frac{h}{2eL^2}$.
  \item “$I_1+I_2$” shown in Fig.\ref{Comparator+DAC} can be noisy and crude as any noise on them will act as a common-mode noise and what finally matters is the circulating current stored in the loop.
  \item As a comparison with classical circuits, to store a field of 2.07fT we would need a LSB current resolution of less than 5pA circulating in a loop of 100$\mu$m radius which is too fine by a long shot for existing converters.
\end{itemize}
This same differential current is made to flow through another square loop at the other end of the loop-filter which basically acts as the feedback, recall Fig. 1 from earlier in this paper. The vertical running lines which carry the currents $I_1$ and $I_2$ to the input side of the loop-filter would not interfere with the existing supercurrent loops due to magnetic field screening from Meissner effect. Thus comes forth another benefit of utilizing the internally circulating supercurrent rings. \par
Also a careful look at the structure would yield that the comparator alongwith the feedback is effectively working as a multibit system without any extra budgeting for randomizing the mismatch in feedback elements. Methods involving the budgeting and removal of mismatch in feedback elements of classical $\Delta\Sigma$-ADCs be it continuous-time or discrete-time consume a huge amount of designer's time, effort not to mention the area and power penalties. This also to some extent limits the speed in the continuous-time versions of these ADCs as compensation of the excess loop delays from randomizing and shaping this mismatch consumes clock time. The step-size of this quantizer is thus given by whatever $\Delta I$ can store $B_{LSB}$ in the loop.

The number of steps in the quantizer is obtained using the same Biot–Savart law as follows
\begin{equation}
B_{LF-max} = \frac{\sqrt{2}\mu_0 I}{\pi L}
\end{equation}
\begin{equation}
B_{LF-min} = \frac{h}{2eL^2}
\end{equation}
\begin{equation}
N_{Lev} = \frac{B_{LF-max}}{B_{LF-min}} = \frac{2\sqrt{2}\mu_0eLI}{\pi h}
\end{equation}
where $I=\frac{I_1+I_2}{2}$. If we have a square loop with each side being 200$\mu$m and $I=9.371mA$ then $N_{Lev}=512$.

\begin{table*}[ht]
\caption{Comparison with existing ADC types published in journals}
\centering
\begin{tabular}{|c|c|c|c|c|}
\hline
Performance Metric & \cite{ADI_CTDSM} & \cite{ADI_DTDSM} & \cite{ADI_SAR} & This Work\\
\hline
Type & Continuous-Time & Discrete-Time & Precision & Using proposed\\
        & $\Delta-\Sigma$ ADC & $\Delta-\Sigma$ADC & SAR ADC & Devices and Circuits\\
\hline
SNDR & 107dB & 105.3dB & 96.2dB & With comparable power numbers can target $\geq$135dB \\
\hline
Bandwidth & 391.5KHz & 125KHz & 175KHz & Can be $\geq$1MHz depending on speed of digital logic \\
\hline
Area\footnote & N/A & N/A & N/A & Will heavily depend on fabrication methods \\
\hline
Power\footnote & 126mW & 34mW & 44mW & $I_{Q}$ of comparator-loop, input-drivers, digital logic \\
\hline
Supplies & 3.3V, 1.8V, 1.1V & 5V, 2V, 1.1V & 5V, 1.8V, 1.1V & Will need just one supply \\
\hline
$\vec{E}$ and $\vec{B}$ shielding & No & No & No & Yes \\
\hline
\multicolumn{4}{c}{$Area^{1}$ Exact die not available, package dimensions reliable metric}\\
\multicolumn{4}{c}{$Power^{2}$ Multiple supply domains for Analog, Digital and I/O blocks}\\
\hline
\end{tabular}
\end{table*}

\section{Conclusion}
In this paper we have presented a device which leverages the macroscopic quantum phenomena of superconductivity to modify the existent building blocks of complex electonic circuits. For the sake of brevity we have jumped detailed derivations to draw conclusions from the final results which give us an indication of the performance metrics that can be extracted and fabrication challenges we might face.

\section*{APPENDIX}
\setcounter{section}{1}
\section*{Appendix-A : Noise calculations}
We will begin by a very brief review of the flicker noise process occuring in MOSFETs. A typical n-type MOSFET is shown in the diagram below.
\begin{figure}[H]
\includegraphics[scale=0.75]{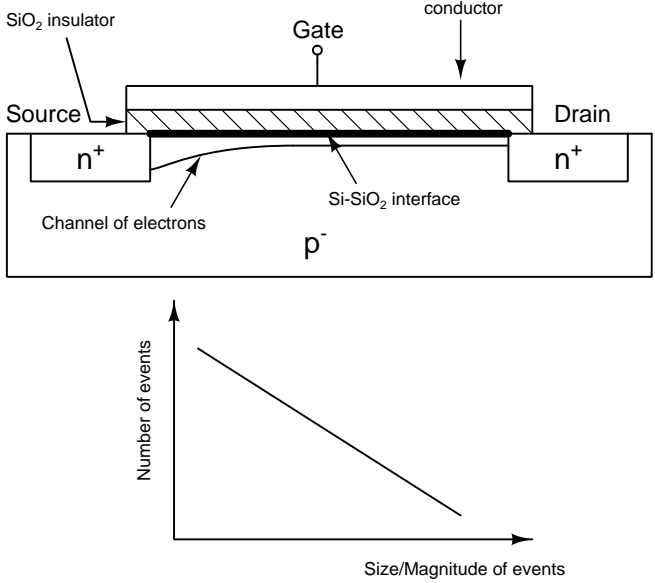}
\caption{A typical n-type MOSFET device showing one of the origins of flicker noise from the traps in silicon and oxide interface. Also shown is a typical variation of a power law phenomenon.}
\end{figure}
An analogy can be drawn from the burst-noise which occurs when a trap or impurity catches a charge carrier and then releases it after some relaxation time $\tau_1$. The auto-correlation function and power spectral density are given by
\numberwithin{equation}{section}
\begin{equation}
R_{XX}(\tau) = R_{XX}(0)e^{-\frac{\vert \tau \vert}{\tau_1}}
\end{equation}
\begin{equation}
S_X(\omega) = \frac{4R_{XX}(0)\tau_1}{1+(\tau_1\omega)^2}
\end{equation}
Flicker noise can be considered as an assortment of many such trapping and releasing events by impurities, lattice defects, interface defects, etc with the last one being the dominant in modern MOSFETs. Also shown in the above diagram is the power law of statistics which in our specific case can be formulated to state that the events with lower disruptive potential occur more than those with higher disruptive potential. This tells us that if we have a very clean semiconductor sample, the bulk of trapping-releasing activites would be from the silicon and oxide interface. These will have lowest relaxation times followed by those caused by lattice defects. Thus we can conclude that the number of such events would be inversely proportional to the magnitude and the relaxation time $\tau_1$. Integrating over all such processes to get the final power spectral density of the flicker noise
\begin{equation}
\begin{split}
S_X(\omega) & = \int_{\tau_1}^{\tau_2} N(\tau) \frac{4R_{XX}(0)\tau}{1+(\tau\omega)^2} d\tau \\
					    & = \int_{\tau_1}^{\tau_2} \frac{k}{\tau} \frac{4R_{XX}(0)\tau}{1+(\tau\omega)^2} d\tau \\
					    & = \frac{k^{'}}{\omega} \int_{\omega \tau_1}^{\omega \tau_2} \frac{d(\omega \tau)}{1+(\tau \omega)^2} \\
					    & = \frac{k^{'}}{\omega} \bigg[ tan^{-1}(\omega \tau_2) -tan^{-1}(\omega \tau_1) \bigg] \\
\end{split}
\end{equation}
This is the typical inverse dependence on frequency seen for flicker noise. In the case of the device proposed in the earlier sections, the trapping and releasing action seen at the oxide interface in MOSFETs is absent as the total cylinder is made up of the same material with some portions in superconducting mode while some in normal mode due to the action of the E-coils. However those due to impurity atoms or lattice imperfections will still be present. Thus overall compared to its classical couterparts, the proposed device will exhibit much lower flicker noise power spectral density. \par
In the case of thermal nosie, we will begin by considering the case of the normal resistor or a metal wire. Here the random but zero-mean ($\mu$ = 0) voltage at the two terminals is primarily due to the scattering of electrons in the three dimensional gas moving in Brownian motion. The scattering is a function of the mean velocity of the electrons and are thus dependent on the temperature.
\begin{figure}[H]
\centering
\includegraphics[scale=0.7]{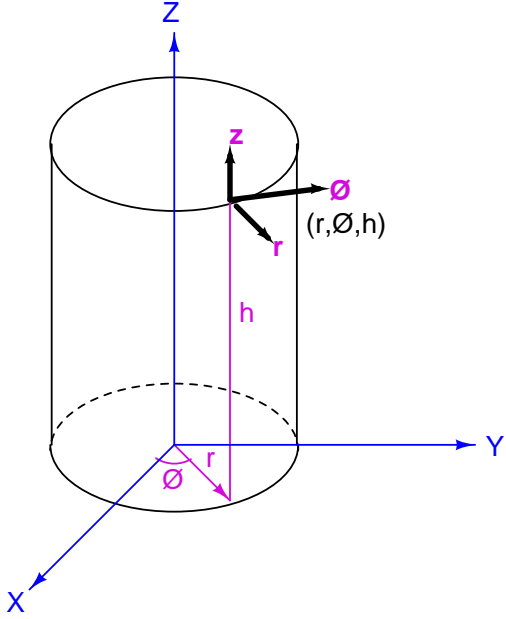}
\caption{Cartesian to cylindrical co-ordinate transformation.}
\end{figure}
Consider an electron travelling towards any one of the terminals of the resistor or wire. If there is a scattering event in any $\vec{x}$ or $\vec{y}$ or $\vec{z}$ directions, then we would notice a change in the voltage measured across the terminals. Or in other words, all the three directions of scattering would lead to an overall noise. \par
In the case of the device proposed, we would shift to the cylindrical coordinates system instead of the cartesian coordinate for ease without any loss of generality. Here since the information is stored in form of the trapped magnetic field due to the circulating supercurrent, any fluctuations in the supercurrent would lead to a noise. Let us consider a Cooper-pair which is travelling along the concentric region inside the superconductor then slight variations in its $\vec{z}$ due to scattering from impurities would not matter. Also any variations in the $\vec{r}$ also would not impact the stored magnetic field. The only scattering direction that will cause a change in the magnetic field would be a variation in $\vec{\phi}$ direction. Thus we can say that only one out of the three degrees of freedom for scattering would impact our stored magnetic field. Or in other words this superconducting device is three times more robust than its classical counterpart MOSFETs. The normal to superconductor tunnelling current $I_{SN}$ would however contribute to additional noise. The spectral characteristics of it is unknown at this point.

\setcounter{section}{2}
\section*{Appendix-B : Errors from traps and impurities}
In this portion we would consider the effects of non-magnetic impurities which are weak disorders and do not cause the eigenstates of the single particle Hamiltonian to be localized in space. Let us consider the Bogoliubov–de Gennes equations -:
\numberwithin{equation}{section}
\begin{equation}
\bigg[-\frac{\hbar^2}{2m}\nabla^2 - \mu + U(\vec{r}) \bigg]u(\vec{r}) + \Delta v(\vec{r}) = \epsilon u(\vec{r})
\end{equation}
and
\begin{equation}
-\bigg[-\frac{\hbar^2}{2m}\nabla^2 - \mu + U(\vec{r}) \bigg]v(\vec{r}) + \Delta^* u(\vec{r}) = \epsilon v(\vec{r})
\end{equation}
Here U($\vec{r}$) describes the electrostatic potential due to the impurity. Now consider a single-electron wavefunction of a normal metal $w_n(\vec{r})$ that satisfies
\begin{equation}
\bigg[-\frac{\hbar^2}{2m}\nabla^2 - \mu + U(\vec{r}) \bigg]w_n(\vec{r}) = \xi_n w_n(\vec{r})
\end{equation}
With $\Delta$ being a constant, we have
\begin{align}
u_n(\vec{r}) = w_n(\vec{r})U_n \\
v_n(\vec{r}) = w_n(\vec{r})V_n
\end{align}
where the coherence factors and the energy spectrum is given by 
\begin{equation}
\vert U_n \vert ^2 = \frac{1}{2}\bigg( 1 + \frac{\xi_n}{\epsilon_n} \bigg)
\end{equation}
\begin{equation}
\vert V_n \vert ^2 = \frac{1}{2}\bigg( 1 - \frac{\xi_n}{\epsilon_n} \bigg)
\end{equation}
\begin{equation}
\epsilon_n= \sqrt{\xi_n^2 + \vert\Delta\vert^2}
\end{equation}
Thus we see that if the elastic scattering mean free path is more than the coherence length inside the superconductor and the desity of states is not significantly affected by the disorder parameter then we reach the same $\Delta$ and $T_C$ as obtained by the BCS theory. \par
However, if we have a magnetic impurity scattering which leads to spin-flips of the Cooper-pair it can result in loss of superelecton density. Thus depending on the impurity atom's size and denisty of such impurities throughout the superconductor the overall supercurrent denisty, $\Delta$ and $T_C$ will be modified.

\setcounter{section}{3}
\section*{Appendix-C : SIN junction based device as another alternative to the proposed device}
Here we will look at another alternative to the device proposed in the main section of this paper. Whereas that works by sliding the supercurrent rings up through the successive turning on/off of the E-coils, the following device works on the principle of sliding the supercurrent rings through the application of a transverse electric field on either ends of the cylinder. Following diagram shows the same cylindrical construction as the original device.
\begin{figure}
\includegraphics[scale=0.75]{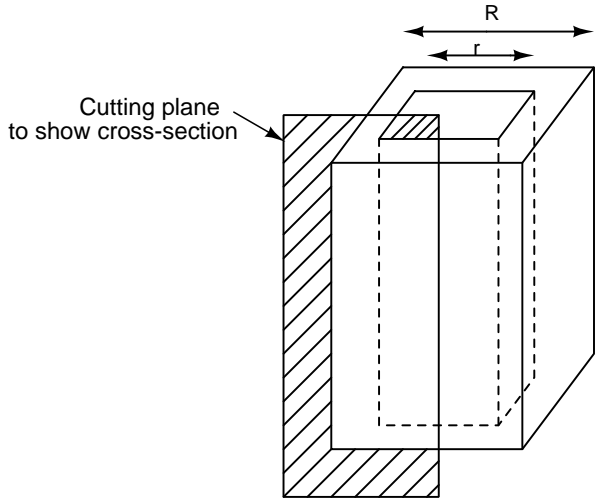}
\caption{Cylindrical construction of the original device.}
\end{figure}
The cutaway section shaded in the above figure is shown below with the new addition of the insulating and conducting layers at the two ends of the cylinder. The working and setup of the magentic field is same as the previous device upto the point where the external magnetic field is frozen inside the cylindrical cavity by the circulating supercurrent bunces spread out over the height of the cylinder. At this point the E-coils are disengaged and an opposite polarity voltage is applied on the top and bottom plates of the structure. Assuming the top plate is applied a potential $+V$ and the bottom plate is applied the opposite potential $-V$, this causes the supercurrent rings which are of negative charge carriers to shift upwards towards the top-plate like a capacitor action with the insulating dielectric between the conducting metal plates.
\begin{figure}
\includegraphics[scale=0.7]{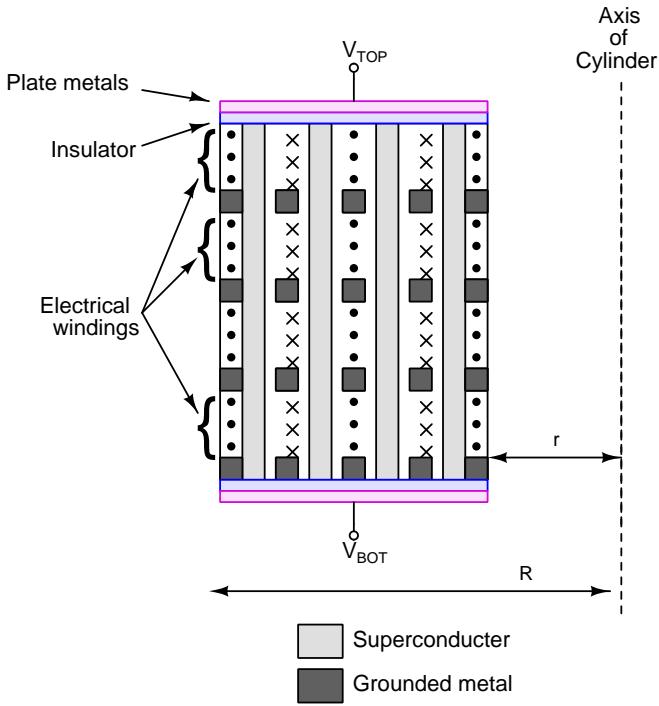}
\caption{Construction of the cutaway section.}
\end{figure}
When the rings converge at the top end of the structure this will lead to an amplification of the frozen magnetic field thus giving us another amplifier topology. This device incorporates another type of junction called Normal-Insulator-Superconductor (NIS) which we will briefly describe here and mention some relevant equations. \par
We again begin from the Bogoliubov–de Gennes equations but modify them based on the following assumptions -:
\begin{itemize}
  \item The normal metal and the superconductor are separated by the insultor extending in the $\hat{y}$ and $\hat{z}$ directions. Thus we have a variation of potential along the $\hat{x}$ direction only. This causes the wavefunction to take the form\\ $e^{i(k_yy+k_zz)}\begin{bmatrix}u(x)\\v(x)\end{bmatrix}$.
  \item The momentum along $\hat{y}$ and $\hat{z}$ directions are conserved, $E_x = E_F - \frac{\hbar^2(k_y^2+k_z^2)}{2m}$
  \item We model the insultating region as a thin layer and the potential barrier by $U(x) = I\delta(x)$ where $I$ represent the height of the potential barrier.
\end{itemize}
Also, here we assume that the incident electron from the normal region will be reflected back with two components given by the normal reflection where we have an similar particle as the electron going back into the metal alongwith an Andreev reflected anti-particle too.
\begin{equation}
\bigg[-\frac{\hbar^2}{2m}\frac{d^2}{dx^2} - E_x + I\delta(x) \bigg]u(x) + \Delta v(x) = \epsilon u(x)
\end{equation}
\begin{equation}
-\bigg[-\frac{\hbar^2}{2m}\frac{d^2}{dx^2} - E_x + I\delta(x) \bigg]v(x) + \Delta^* u(x) = \epsilon v(x)
\end{equation}
Boundary conditions require the solution to $u$ and $v$ to be equal on either side of the insultating barrier. Also the difference between the $\hat{x}$-derivative of $u$ and $v$ on the right and left side respectively should be equal to $\frac{2mI}{\hbar^2}$ times the spatial solution of $u$ and $v$ at $x=0$.
\begin{figure}[H]
\centering
\includegraphics[scale=1.1]{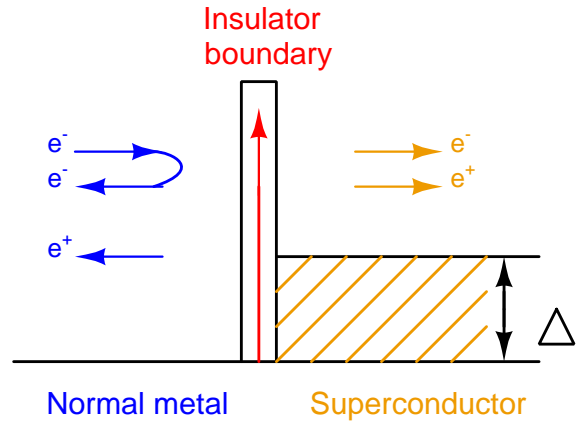}
\caption{A normal metal - insulator - superconductor junction.}
\end{figure}
Consider the case when the energy of the incident electron is more than the energy-gap in the superconductor $\epsilon>\Delta$ which can lead to four possibilities
\begin{itemize}
  \item The resulting transmitted and reflected wavefunctions when an electron is incident on the junction from the normal metal region
  \item The resulting transmitted and reflected wavefunctions when a hole is incident on the junction from the normal metal region
  \item The wavefunctions when an electron is incident on the junction from the superconductor region
  \item The wavefunctions when a hole is incident on the junction from the superconductor region
\end{itemize}
Skipping detailed derivations, we simply state that in the $1^{st}$-case the reflected wavefunction contains an electron and its anti-particle, while an electron and it's anti-particle is transmitted into the superconductor. Similarly in the $2^{nd}$-case the reflected wavefunction contains a hole and also its anti-particle, while an electron and it's anti-particle is transmitted into the superconductor. For the $3^{rd}$ and $4^{th}$ cases we have a similar scenario as in the previous two cases; transmitted electrons and holes alongwith two reflected components of similar and opposite nature to that of the incident particle. \par
Now for the case when $\epsilon<\Delta$ we consider the $1^{st}$-case as an example where a particle is incident on the junction from the normal-metal side. Here the incident particle is reflected as a combination of a particle of same nature and its anti-particle. In the absence of the insulating barrier we would have a reflection of only the anti-particle of the incident particle. \par
To derive the tunneling current across the NIS junction, we consider the case when a voltage V is applied across the interface with the normal metal being at the higher potential. The total tunnelling current is given by the combined contribution from the 4 possible tunnelling scenarios discussed above. Ignoring any charge buildup in the insulator, we can conclude that the current on the normal metal side would be equal to that on the superconductor side. This would be a function of the occupied density of states on the metal side for the incident paticle and the vacant states of its reflected anti-particle on the same side. Skipping the thorough derivations, we arrive at
\begin{equation}
\begin{split}
I_{NIS} &= AeN(0)v_FS \int_{-\infty}^{\infty} \big[1 - \vert b\vert^2 + \vert a\vert^2 \big] \times \big[f_0(\epsilon-eV) \\
&-f_0(\epsilon) \big] d\epsilon
\end{split}
\end{equation}
where A is a constant dependant on the junction's geometry, N(0) is density of states at Fermi surface, $v_F$ is the Fermi velocity, S is the surface area of the junction, $(1 - \vert b\vert^2 + \vert a\vert^2)$ is proportional to the transmission coefficient at the interface and 
\begin{equation}
f_0=\frac{1}{e^{\frac{\epsilon}{T}}+1}
\end{equation}
For low temperatures this can be approximated as
\begin{equation}
I_{NIS} = \frac{Ae^2N(0)v_FS}{1+\big(\frac{mI}{\hbar^2\vert k_x\vert}\big)^2}\sqrt{(eV)^2 - \vert \Delta \vert^2}  \Theta \bigg( V - \frac{\vert \Delta \vert}{e} \bigg)
\end{equation}

\section*{Acknowledgment}
The authors would like to thank... for their valuable inputs.

\bibliographystyle{IEEEtran}

\end{document}